\documentclass{article}
\usepackage{amssymb}

\begin{document}

\title{On the behavior of bosonic systems in the presence of topology
fluctuations}
\author{A. A. Kirillov, E.P. Savelova \\
%EndAName
\emph{Branch of Uljanovsk State University in Dimitrovgrad, }\\
\emph{Dimitrova str 4.,} \emph{Dimitrovgrad, 433507, Russia} }
\date{}
\maketitle

\begin{abstract}
The behavior of bosonic systems in the presence of space-time foam is
analyzed within the simplistic model of a set of scalar fields on a flat
background. We discuss the formula for the path integral which allows to
account for the all possible topologies of spacetime. We show that the
proper path integral originates from the parastatistics suggested first by
H.S. Green and that it defines a cutoff for the field theory.
\end{abstract}

\newpage

\section{Introduction}

Spacetime foam is commonly believed to cure divergencies in particle physics
\cite{wheeler} and therefore it will eventually allow to remove the
unnatural and non-physical (and extremely restrictive) principle of the
renormalizability of physical field theories\footnote{%
In particular, general relativity itself represents a non-renormalizable
theory.}. However so far such a property has not been explicitly established
yet.

At first glance the basic difficulty here stems from the problem of
classifying topologies in 4-dimensions. Indeed, the adequate description of
spacetime foam effects is reached in the euclidean quantum gravity advocated
primary by S. W. Hawking \cite{H78} and developed by many authors (e.g., see
Refs. \cite{bab}-\cite{Banks}). The euclidean path integral \ for the
expectation value of an observable $B$ is
\begin{equation}
\left\langle B\right\rangle =\frac{\sum Be^{-S}}{\sum e^{-S}}  \label{z1}
\end{equation}%
where $S$ is the euclidean action and sum is taken over all field
configurations and all topologies of the euclidean spacetime. The path
integral is usually supposed to be taken in the two steps. First, one
integrates over all field configurations keeping a specific topology fixed
and then sums over different topologies, so that the partition function can
be presented as%
\begin{equation}
Z=\sum e^{-S}=\sum_{topologies}e^{-S_{eff}}  \label{z}
\end{equation}%
where $S_{eff}$ is an independent effective action for each topology. Now
one may use the semiclassical approximation (instantons) to evaluate
contributions of different topological classes etc. and this is the way on
which the further development of euclidean quantum gravity is going on
(e.g., see Refs. \cite{hwu} and references therein). We leave aside the loop
quantum gravity \cite{loop}, for essentials remain the same (as far as
topologies is concerned).

It is clear however that results obtained on this way are rather restrictive
in nature. Save the absence of an appropriate classification of different
topologies, one can never justify that terms (topological classes) omitted
give small effects. Even if such terms have bigger actions $S_{eff}$ the
number of such additional terms is enormous. One may think that the
semiclassical approximation in (\ref{z}) (though useful in investigating
particular features) is not suitable. And indeed, if we believe that quantum
gravity (i.e., topology fluctuations) provide quantum fields with a cutoff,
then at very small scales (i.e., at very high energies) the physical space
is effectively absent and all particular topologies may give only a
negligible contribution to (\ref{z}); for every term in (\ref{z})
corresponds to a divergent quantum theory and as we shall see latter on
(e.g., see (\ref{zf})) smooth particular topologies are suppressed indeed.

In the present Letter we suggest absolutely equivalent formula for the path
integral which allows to account for the all possible topologies of
spacetime (even those which are apparently not smooth). As we shall see the
proper path integral automatically defines a Lorentz invariant cutoff for
the field theory as it was to be expected. The formula suggested follows
quite naturally from the three well-established fundamental facts. 1) Any
4-dimensional manifold can be continued to the whole Euclidean space by
adding non-physical regions of the space. Such a continuation is not unique
however. In particular, the existence of a universal covering is the
well-known mathematical fact. However in the general case the universal
covering requires considering a curved space, while at high energies (at
least at laboratory scales) the space looks to be flat. Our claim is that
there always exists a continuation when the space remains to be flat (e.g.,
see Ref. \cite{KS07}). 2) The discrepancy between the actual Green functions
and those for the euclidean space is described by a topological bias of
sources (i.e., the topology or the proper boundary conditions for the actual
Green functions can be accounted for by additional sources). 3) The
topological bias of sources has an equivalent description in terms of
multi-valued fields. We stress that it is the basic fact which allows us to
reformulate the sum over topologies in terms of the sum over multi-valued
field configurations.

The first two facts represent the well known classical results. E.g., the
universal covering (which is not more than the astrophysical way of the
extrapolation of the laboratory coordinate system) and the concept of the
topological bias were described in detail in Refs. \cite{KS07,KT07}. In
particular, in astrophysics when we look at the sky we always have deal with
the universal covering and this allows to give the most natural explanation
for the all the variety of the observed dark matter phenomena (see the above
papers and Ref. \cite{KT06} where theoretical rotation curves for spiral
galaxies were shown to be in a very good agreement with observations). The
bias of sources and the "standard" continuation (i.e., without introducing a
non-flat metric) is the standard tool for solving different electrostatic
problems in classical electrodynamics (e.g., see the image method in Ref.
\cite{jackson}).

The last fact (the multi-valued nature of fields) being transparent is
however less known. The basic construction was suggested in Ref. \cite{K99}
and developed in Refs. \cite{KT02,K03}. We stress that the fact that any
particular topology admits an equivalent description in terms of
multi-valued fields was proved in Ref. \cite{K99}. It turns out that
multi-valued fields have the realization in terms of the so-called
generalized statistics suggested first by H.S. Green in the attempt to solve
the problem of mathematical inconsistencies (renormalization and
regularization procedures) in quantum field theories. We also point out that
in fact multi-valued fields represent the most natural tool to describe the
so-called coda waves and seismic noise \cite{coda}. Indeed, due to multiple
scattering on topology (or in porous systems on boundaries) plane waves are
not solutions to linear field equations (for a particular topology the
homogeneity of space is broken\footnote{%
The homogeneity holds only for mean statistical values.}). Thus if we
consider any wave packet $\phi _{0}$ it, due to multiple scattering,
transforms to $\phi =\sum \phi _{j}$. When the topology is random, the
scattering randomizes phases and such a field acquires the diffuse nature $%
\left\langle \phi ^{2}\right\rangle =\sum \left\langle \phi
_{j}^{2}\right\rangle $, i.e., each term can be considered as an independent
field. Thus although on the micro scale the field equations remain unchanged
the intensities follow a diffusion equation. The diffuse nature of seismic
fields has intensively been studied (e.g., see Refs. \cite{dif} and
references therein). We point out that the physical field (which is measured
in experiments) represents only the sum of terms $\phi =\sum \phi _{j}$ and
it is defined only in the physically admissible region of space. Every term
however becomes an "independent field" upon a continuation to the whole
coordinate space. In quantum theory particles which are described by the
diffuse fields obey the generalized statistics (in particular, the violation
of the Pauli principle in such fields has a rather clear physical sense; the
violation occurs due to the existence of "mirror" particles in non-physical
regions of space, while upon restriction to the fundamental domain the
statistics restores) \cite{K03}.

For the sake of simplicity and to make the basic ideas clear (and to avoid
usual technical problems in quantum gravity) we, in the present paper,
consider the most simple example of a set of scalar fields in $R^{4}$. The
metric is supposed to be everywhere flat, while the topology is described by
some gluing procedure along some multi-connected hypersurfaces. We point out
that in general when considering the universal covering such gluing leads to
$\delta $- like singularities in the scalar curvature which rigorously
speaking require to account for the gravitational action. To avoid such
problem we shall suppose that every hypersurface is approximated by
piecewise flat surfaces. Then the $\delta $- like terms in the curvature are
concentrated on \ vertexes and ribs which have zero measure and do not
contribute to the geodesic flow. Moreover such terms possess both signs
(depending on the induced curvature on the hypersurfaces) and for
sufficiently complex topologies the vanishing of the mean curvature is
actually not restrictive. In considering the standard continuation (by the
image method) the metric remains everywhere flat, while the scattering on
the topology is completely described by the bias of sources and we need not
to add the gravitational action.

\section{The universal covering and the topological bias}

The universal covering for an arbitrary non-trivial topology of space can be
constructed as follows. We take a point $O$ in our space $\mathcal{M}$ \ and
issue geodesics (straight lines) from $O$ in every direction. Then points in
$\mathcal{M}$ can be labeled by the distance from $O$ and by the direction
of the corresponding geodesic. In other words, for an observer at $O$ the
space $\mathcal{M}$ will always look as $R^{4}$. However if we take a point $%
P\in \mathcal{M}$, there may exist many homotopically non-equivalent
geodesics connecting $O$ and $P$. Thus, any source at the point $P$ will
have many images in $R^{4}$. The topology of $\mathcal{M}$ can be determined
by noticing that in the observed space $R^{4}$ there is a fundamental domain
$\mathcal{D}$ such that every point in $\mathcal{D}$ has a number of copies
outside $\mathcal{D}$. The actual manifold $\mathcal{M}$ is then obtained by
identifying the copies. In this way, we may describe the topology of space $%
\mathcal{M}$ by indicating for each point $r\in R^{4}$ the set of its copies
$E(r)$, i.e. the set of points that are images of the same point in $%
\mathcal{M}$.

Consider now the actual Green function for a scalar wave equation in the
physically admissible region $\mathcal{D}$
\[
\left( -\square _{x}+m^{2}\right) G\left( x,y\right) =4\pi \delta \left(
x-y\right) ,
\]%
where $x$, $y\in \mathcal{D}$. Upon continuing to the universal covering $%
R^{4}$ this equation transforms as follows%
\begin{equation}
\left( -\square _{x}+m^{2}\right) G\left( x,y\right) =4\pi N\left(
x,y\right) ,  \label{gr}
\end{equation}%
where coordinates $x$, $y$ are extended to the whole space $R^{4}$ and
\begin{equation}
N(x,y)=\delta \left( x-y\right) +\sum \delta \left( x-f_{i}(y)\right)
\label{b}
\end{equation}%
(the sum is here taken over all images of the point $y$, i.e.,\ over all $%
f_{i}(y)\in E(y)$). The two point function $N\left( x,y\right) $ was called
the topological bias in Refs. \cite{KT07,K06} which describes the
discrepancy between the actual physical space (the fundamental domain $%
\mathcal{D}$) and the universal covering (the simple topology space) $R^{4}$%
. We point out that the topology is completely (one-to-one) defined by the
specifying the bias $N\left( x,y\right) $ (\ref{b}). In quantum gravity
(when topology may fluctuate) the bias becomes an operator valued function
which has the meaning of the density of extra images for the actual source $%
\delta \left( x-y\right) $.

The structure of the bias (\ref{b}) on the universal covering has one
important feature which allows it to mimic dark matter phenomena (which are
discussed in detail in Refs. \cite{KT06,KT07}), i.e.,
\[
\int_{V}N(x,y)d^{4}x=N\left( V\right) \geq 1
\]%
where $V$ is some volume around the point $y$. The number $N\left( V\right)
-1=0,1,2,...$ gives the number of points $f_{i}(y)$ which get into the
coordinate volume $V$. Roughly, this number characterizes how many times the
volume $V$ covers the fundamental domain (or the physically admissible
region) $\mathcal{D}$.

Let us return to the path integral (\ref{z}). Consider a particular virtual
topology of space. It is clear that the action in (\ref{z1})-(\ref{z}) has
the same value for all physical spaces which can be obtained by rotations
and transitions of the coordinate system in $R^{4}$. Thus, upon averaging
out over possible orientations and transitions the bias acquires always the
structure $N\left( x,y\right) =N\left( \left\vert x-y\right\vert \right) $
and for the Green function we find%
\begin{equation}
G\left( x-y\right) =\int \frac{d^{4}k}{\left( 2\pi \right) ^{4}}\frac{%
N\left( k\right) }{k^{2}+m^{2}}\exp \{ik\left( x-y\right) \},  \label{fg}
\end{equation}%
where $N\left( k\right) $ is the Fourier transform for the bias. The above
Green function plays the most important role in particle theory and its UV
(ultra-violet) behavior (actually that of the bias $N\left( k\right) $)
defines whether the resulting quantum theory is finite or not. What we
expect that the proper definition of the path integral over virtual
topologies should fix the specific form of the bias $N\left( k\right) $.

We also point out that the universal covering is what we actually use in
astrophysics when extrapolating our laboratory coordinate system to
extremely large distances. Therefore, in expressions (\ref{b}) (\ref{fg})
the coordinates $x$, $y$ have the direct physical (observational) status in
applying to cosmological problems (DM and dark energy phenomena, origin of
density perturbations etc.). In particular, we can never say (without
additional subtle effects) if two points $x_{1}$ and $x_{2}$ are close or
not (at least there are no external safe rulers to measure the distances).
On the contrary, in high energy physics we use an extrapolation to very
small scales (by means of our "safe" laboratory rulers). Again we cannot say
if two points $x_{1}$ and $x_{2}$ are close or not. However we still can
assign specific distances extrapolated from the laboratory coordinate system
and this is exactly the coordinate system we use in particle physics. As we
shall see the extrapolation in particle physics leads to the same
expressions (\ref{gr}), (\ref{fg}) however the bias (\ref{b}) acquires
somewhat different features\footnote{%
As it was shown in Ref. \cite{K06} in this case the bias $N\left( x,y\right)
$ represents a projection operator onto physically admissible states. This
means that $(\widehat{N})^{2}=\widehat{N}$ and in the basis of eigenvectors
it takes the form $N\left( x,y\right) $ $=$ $\sum N_{k}f_{k}^{\ast }\left(
x\right) f_{k}\left( y\right) $ with eigenvalues $N_{k}=0,1$. While on the
universal covering possible eigenvalues $N_{k}=0,1,2,...$.}. By other words
the Universe looks somewhat different when we look at small or large
distances.

\section{Topological bias in particle physics}

In the present section we consider the bias which originates from a single
wormhole. Such a bias was constructed first in Ref. \cite{KS07} for the
massless field in 3-dimensions, while the generalization to the euclidean
4-space is straightforward. We point out that a wormhole describes a virtual
baby universe which may branch off and joint onto our mother Universe \cite%
{bab}-\cite{Banks}.

A single wormhole can be viewed as a couple of conjugated spheres $S_{\pm }$
of the radius $a$ and with a distance $d=\left\vert \vec{R}_{+}-\vec{R}%
_{-}\right\vert $ between centers of spheres. The interior of the spheres is
removed and surfaces are glued together. For the sake of simplicity we
consider the massless case i.e., the Green function $\triangle G(x,y)=4\pi
\delta (x-y)$ for such a topology. In Ref. \cite{KS07} we have shown that
the proper boundary conditions (the actual topology) can be accounted for by
adding the bias of the source%
\[
\delta (x-y)\rightarrow \delta (x-y)~+b\left( x,y\right)
\]%
where in the approximation $a/d\ll 1$ the bias in $R^{3}$ takes the form
\begin{equation}
b\left( x\right) \approx a\left( \frac{1}{R_{-}}-\frac{1}{R_{+}}\right) %
\left[ \delta (\vec{x}-\vec{R}_{+})-\delta (\vec{x}-\vec{R}_{-})\right]
\end{equation}%
where we set $y=0$ and neglect the throat size, i.e., all additional sources
(ghost images) are placed in the centers of spheres. The generalization to
the space $R^{4}$ is trivial and gives%
\begin{equation}
b\left( x\right) =a^{2}\left( \frac{1}{R_{-}^{2}}-\frac{1}{R_{+}^{2}}\right) %
\left[ \delta (\vec{x}-\vec{R}_{+})-\delta (\vec{x}-\vec{R}_{-})\right] .
\label{b1}
\end{equation}%
We see that unlike (\ref{b}) the function $b(x)$ has the property $\int
b(x)d^{4}x=0$ which gives $\int N(x)d^{4}x\equiv 1$ and for any volume $V$
we get $N\left( V\right) \leq 1$.

Let us introduce the probability distribution for parameters of the wormhole
$P\left( R_{\pm },a\right) $ which is defined by the action in (\ref{z1}).
It is clear that due to homogeneity an isotropy of $R^{4}$ this function may
depend only on $d=\left\vert \vec{R}_{+}-\vec{R}_{-}\right\vert $ and we
find for the mean bias
\begin{equation}
\overline{b}\left( r\right) =2\int \left( \frac{1}{R^{2}}-\frac{1}{r^{2}}%
\right) f\left( \left\vert \vec{R}-\vec{r}\right\vert \right) d^{4}\vec{R},
\label{b2}
\end{equation}%
where $f\left( d\right) $ $=$ $\int a^{2}P\left( d,a\right) da$. For the
Fourier transforms $b\left( k\right) =\left( 2\pi \right) ^{-2}\int b\left(
r\right) $ $e^{-ikr}d^{4}r$ this expression takes the simplest form
\begin{equation}
\overline{b}\left( k\right) =\frac{8\pi \left( f\left( k\right) -f\left(
0\right) \right) }{k^{2}}.  \label{b_k}
\end{equation}

In the so-called long-wave approximation (the low energy physics) we can
completely neglect the throat size $a\rightarrow 0$. In this limit the
action for the wormhole does not depend on the separation distance $%
d=\left\vert \vec{R}_{+}-\vec{R}_{-}\right\vert $ at all, i.e., $P\left(
d,a\right) =P\left( a\right) $, and the mean bias reduces merely to $%
\overline{b}\left( x\right) =b\delta \left( x\right) $ (e.g., see Ref. \cite%
{KS07}). Therefore, the effect of wormholes reduces merely to a
renormalization of physical constants (e.g., of charge values) which is in
the complete agreement with the previous results of Refs. \cite{Col1, Col2,
Banks}. Moreover, the value $b<0$ \cite{KS07} which means that virtual
wormholes always diminish charge values as it was first pointed out in Ref.
\cite{Banks}.

In conclusion of this section we point out that the multiplier $4\pi /k^{2}$
in (\ref{b_k}) and $1/R_{\pm }^{2}$ in (\ref{b2}) is the standard Green
function for $R^{4}$. In the case of massive particles it should be replaced
with $4\pi /\left( k^{2}+m^{2}\right) $ and $-\frac{m^{2}}{8\pi z}%
H_{1}^{\left( 2\right) }\left( z\right) $ (where $H_{1}^{\left( 2\right)
}\left( z\right) $ is the second order Hunkel function and $%
z^{2}=-m^{2}R_{\pm }^{2}$) respectively. Thus, we see that in particle
physics the structure of the Green functions (\ref{gr}), (\ref{fg}) remains
the same, while the property of the bias for the universal covering $N\left(
V\right) \geq 1$ changes drastically to $N\left( V\right) \leq 1$. This
feature reflects the two possible different ways of the continuation of the
physical space $\mathcal{M}$ to the coordinate space $R^{4}$ (e.g., see
discussions in Ref. \cite{KS07}). We recall that in particle physics the
bias $N\left( x,y\right) $ can be considered as a projection operator onto
physically admissible states (e.g., see Sec.2 in Ref. \cite{K06}), which
means that it always has eigenvalues $N_{i}=0,1$.

\section{Multi-valued fields and generalized statistics}

The structure of the bias (\ref{b}) suggests the analogous decomposition of
the true Green functions
\begin{equation}
G\left( x,y\right) =G_{0}\left( x-y\right) +\sum G_{0}\left(
x-f_{i}(y)\right)  \label{gr2}
\end{equation}%
where $G_{0}\left( x-y\right) $ is the standard Green function for the
euclidean space $R^{4}$. If we present it in the form of the path integral
for a scalar particle in $\mathcal{D}$, i.e., $G\left( x,y\right) $ $%
=\sum_{x\left( s\right) \in \mathcal{D}}\exp \left( -\int_{y}^{x}ds\right) $
then every term in (\ref{gr2}) corresponds to the restriction of
trajectories $x\left( s\right) $ to a particular homotopic class \cite{KT07}%
. When we continue such terms to the whole space $R^{4}$ they acquire the
character of independent fields that is to say that such particles has to be
described by a scalar field $\phi $ which acquires the multi-valued
(diffused) nature. An equivalen representation for such a field can be
achieved in terms of the generalized statistics (e.g., see for details Ref.
\cite{K03}). For the sake of convenience we present in the present section
basic ellements of the generalized statistics and generalized second
quantization suggested first by H.S. Green. It is remarkable that the basic
motivation for the generalized quantization method used by H.S. Green was
the presence of mathematical inconsistencies (renormalization and
regularization procedures) in quantum field theories. In the present paper
we demonstrate that the goal (the removal of the inconsistencies) is
actually reached.

Consider a system of identical particles with an undefined a priori symmetry
of wave functions. We shall use the Bogoliubov's method \cite{B}, in which
the second quantization is applied to the density matrix (to the case of
para-statistics this approach was extended in Ref. \cite{gov}). Let us
define operators $M_{ij}$ of transitions for particles from a quantum state $%
j$ into a state $i$. These operators must obey the Hermitian conditions,
i.e.
\begin{equation}
M_{ik}^{+}=M_{ki}
\end{equation}%
and the algebra $SU\left( N\rightarrow \infty \right) $ ($N$ is the number
of different one--particle quantum states)
\begin{equation}
\left[ M_{ij},M_{km}\right] =\delta _{jk}M_{im}-\delta _{im}M_{kj}.
\label{NB}
\end{equation}%
These conditions give the algebraic expression of the indistinguishability
principle for identical particles. For Bose and Fermi statistics they first
were established by N.N. Bogoliubov \cite{B} and generalized to the case of
arbitrary statistics by A.B. Govorkov \cite{gov1}.

Consider now systems with a variable number of particles. To this end we
need to consider a set of creation and annihilation operators for particles (%
$a_{i}^{+}$ and $a_{k}$) and somehow express via them the transition
operators $N_{ij}$. The simplest generalization of Bose and Fermi statistics
was first suggested by H.S. Green \cite{G53} and latter by D.V. Volkov \cite%
{Vol} and are called the parastatistics or the Green-Volkov statistics.

Consider a set of creation and annihilation operators of particles $%
a_{i}^{+} $ and $a_{k}$, while the transition operators we present in the
form
\begin{equation}
M_{ik}=\frac{1}{2}\left( a_{i}^{+}a_{k}\pm a_{k}a_{i}^{+}\mp N_{ik}\right) ,
\label{Nd}
\end{equation}%
where $N_{ik}$ is, in general, an arbitrary Hermitian matrix. The upper sign
stands for the generalized Bose statistics, while the lower sign stands for
the Fermi statistics. The operator $M_{i}=M_{ii}$ has sense of the particle
number operator in the quantum state $i$. Then the creation and annihilation
operators should obey the requirements
\begin{equation}
\left[ M_{i},a_{k}\right] =-\delta _{ik}a_{k},\ ~~\ \ \left[ M_{i},a_{k}^{+}%
\right] =\delta _{ik}a_{k}^{+}.  \label{NA}
\end{equation}%
Consider now a unitary transformation%
\begin{equation}
a_{i}^{\prime }=\sum_{k}u_{ik}a_{k},\ \ \ \ \ a_{i}^{\prime
+}=\sum_{k}u_{ik}^{\ast }a_{k}^{+},
\end{equation}%
where $\sum_{m}u_{im}u_{km}^{\ast }=\delta _{ik}$, under which the relations
(\ref{NB}) and (\ref{NA}) remain invariant. Then applying to (\ref{NA}) an
infinitesimal transformation $u_{ik}=\delta _{ik}+\varepsilon _{ik}$, $%
\varepsilon _{ik}^{\ast }=-\varepsilon _{ki}$ and retaining the first order
terms in $\varepsilon $ we get the basic commutation relations for the
creation and annihilation operators%
\begin{equation}
\lbrack M_{kl},a_{m}^{+}]=\delta _{lm}a_{k}^{+},\ \ \ \ \
[M_{lk},a_{m}]=-\delta _{lm}a_{k},  \label{par-rel}
\end{equation}%
which were first suggested by Green \cite{G53}.

Consider now the vacuum state $\left\vert 0\right\rangle $ that is
\begin{equation}
a_{k}\left\vert 0\right\rangle =0
\end{equation}%
for all $k$. Then the requirement that for all $i$ and $k$ the transition
operators annihilate the vacuum state
\begin{equation}
M_{ik}\left\vert 0\right\rangle =0
\end{equation}%
leads to the condition on one-particle quantum states in the form
\begin{equation}
a_{k}a_{i}^{+}\left\vert 0\right\rangle =N_{ik}\left\vert 0\right\rangle ,
\label{1st}
\end{equation}%
which means that the basis of one-particle states is, in general, not
orthonormal but has norms $\left\langle 0\right\vert
a_{k}a_{i}^{+}\left\vert 0\right\rangle =N_{ik}$. From the physical
standpoint this signals up the presence of some degeneracy of quantum states
(e.g., the presence of an extra coordinate etc., see discussions in Refs.
\cite{K03}).

The fact that $N_{ik}$ is a Hermitian matrix denotes that there exists a
basis of one-particle wave functions in terms of which this matrix has the
diagonal form, i.e., $N_{ik}=\delta _{ik}N_{k}$. H.S. Green (and after him
all other invesigators) imposed the Lorentz invariance on the form of Eq. (%
\ref{1st}) which imediately gives the simplest case $N_{k}=N$ (where $N$ is
a constant), i.e., the form $N_{ik}=\delta _{ik}N$ remains invariant in an
arbitrary basis\footnote{%
It is easy to see the analogy of the above matrix $N_{ik}$ with the bias $%
N\left( k,k^{\prime }\right) $ introduced previously. This analogy indicates
the existence of a very deep relation between these two operators. We point
out that the bias reflects the discrepancy between the topology of the
actual physical space and that of $R^{4}$. Therefore the Lorentz invariance
imposed on Eq. (\ref{1st}) immediately kills the baby. Since any particular
topology breaks the Lorentz invariance, it may hold only for mean values.}.

The condition that norms of vectors in the Fock space are positively defined
leads to the requirement that $N$ is an integer number which characterizes
the rank of the statistics or the rate of the degeneracy of quantum states
\cite{G53,gov}. In the simplest case the number $N$ corresponds to the
maximal number of particles which admit an antisymmetric (for parabosons and
symmetric for parafermions) state. The case $N=0$ corresponds to the absence
of fields. The case $N=1$ corresponds to the standard Bose and Fermi
statistics.

For the case of a constant rank Green also presented an ansatz which
resolves the relations (\ref{par-rel}) and (\ref{1st}) in terms of the
standard Bose and Fermi creation and annihilation operators
\begin{equation}
a_{p}^{+}=\sum_{\alpha =1}^{N}b_{p}^{(\alpha )+},\ \ \ a_{k}=\sum_{\alpha
=1}^{N}b_{k}^{(\alpha )},  \label{G-r}
\end{equation}%
where $b_{p}^{(\alpha )}$ and $b_{p}^{(\beta )+}$ are the standard Bose
(Fermi) operators as $\alpha =\beta $ ($\left[ b_{p}^{\left( \alpha \right)
}b_{k}^{\left( \alpha \right) +}\right] _{\pm }$ $=$ $\delta _{pk}$) but
anti-commutate (commutate) as $\alpha \neq \beta $ ($\left[ b_{p}^{\left(
\alpha \right) }b_{k}^{\left( \beta \right) +}\right] _{\mp }=0$) for the
case of parabose (parafermi) statistics. The presence of an additional index
$\alpha $ in the creation and annihilation operators removes the
degeneration of one-particle quantum states pointed out.

The Green representation (\ref{G-r}) can be easily generalized to the more
general case of an arbitrary Hermitian matrix $N_{ik}$. In the basis in
which this matrix takes the diagonal form $N_{ik}=N_{k}\delta _{ik}$ the
Green representation is given by the same expression (\ref{G-r}) in which,
however, the rank of statistics $N_{k}$ depends on the quantum state (the
index $k$). Thus, in the general case the rank of statistics represents an
additional quantum variable. We also note that in an arbitrary basis the
Green representation does not work and, therefore, we can say that the
matrix $N_{ik}$ distinguishes a preferred basis of quantum states, see also
discussions in Refs. \cite{K03}.

\section{Multi-valued fields and the action}

In the case of a homogeneous and isotropic topological structure the
multi-valued character of the scalar field is more convenient to describe in
the Fourier representation ($\phi =\frac{1}{\left( 2\pi \right) ^{2}}\int
d^{4}k\phi _{k}e^{ikx}$) that is to replace the single-valued field $\phi
_{k}$ with a set of fields $\phi _{k}^{j}$ where $j=1,2,...N\left( k\right) $%
, while the bias $N\left( k\right) $ has the meaning of the number of such
fields (or the rank of statistics). We recall that from the phenomenological
standpoint such fields were introduced first in Ref. \cite{K99} and for the
relation to the generalized statistics see Refs. \cite{K03}.

Consider now the euclidean action for the scalar field (we use the Planckian
units in which $M_{pl}=1$)%
\begin{equation}
S=\frac{1}{2}\int \left[ \left( \partial _{\mu }\phi \right) ^{2}+m^{2}\phi
^{2}+V\left( \phi \right) \right] d^{4}x.  \label{ac}
\end{equation}%
Rigorously speaking the integral here should run only over the fundamental
domain $\mathcal{D}$. However to describe different possible topologies on
an equal footing we should continue this expression on the whole space $%
R^{4} $. In what follows we shall use the Fourier transform for the field,
while the actual topology will be encoded by specifying $N\left( k\right) $
(we assume that the integration over transitions and orientations in (\ref{z}%
) is already carried out and therefore $N\left( k\right) $ defines a whole
class of topologies, while $S$ is the modified action). Then the linear part
of the action takes the structure\footnote{%
We point out that such a simple form for the linear part of the action is
reached only for isotropic and homogeneous class of topologies, while for a
particular topology the bias has the structure $N=N\left( k,k^{\prime
}\right) $ and the action diagonalizes in a specific (for given topology)
basis e.g., see discussions in Refs.\cite{K03,K06}.}%
\begin{equation}
S_{0}=\frac{L^{4}}{2}\int \sum_{j=1}^{N\left( k\right) }\left(
k^{2}+m^{2}\right) \left\vert \phi _{k}^{j}\right\vert ^{2}\frac{d^{4}k}{%
\left( 2\pi \right) ^{4}},  \label{act}
\end{equation}%
while the non-linear term $S_{int}\left( \phi \right) $ should be accounted
for by perturbations. We recall that in this expression the values of the
number of fields $N\left( k\right) $ depend on scales under consideration
and, therefore, the result for the cutoff function depends on the choice of
the continuation used. As it was explained previously in astrophysical
problems we use the universal covering and the number of fields takes values
$N\left( k\right) $ $=$ $0,1,2,...$, while in particle physics the number of
fields can take only two possible values $N\left( k\right) =0,1$.

The physical sense has the sum of fields, and therefore the generating
functional should be taken as%
\begin{eqnarray}
\widetilde{Z}\left[ J\right] &=&\exp \left\{ -S_{int}\left( \frac{\delta }{%
\delta J}\right) \right\} \int D\left[ \phi \right] \exp \left\{
-S_{0}\left( \phi \right) +L^{4}\int J\left( -k\right) \widetilde{\phi }%
_{k}d^{4}k\right\}  \nonumber \\
&=&\widetilde{Z}\left[ 0\right] \exp \left\{ -S_{int}\left( \frac{\delta }{%
\delta J}\right) \right\} \exp \left\{ \frac{L^{4}}{2}\int \frac{\left\vert
J\left( k\right) \right\vert ^{2}}{k^{2}+m^{2}}N\left( k\right) \frac{d^{4}k%
}{\left( 2\pi \right) ^{4}}\right\}  \label{gf}
\end{eqnarray}%
where $\widetilde{\phi }_{k}=\sum_{j=1}^{N\left( k\right) }\phi _{k}^{j}$,
while for $\widetilde{Z}\left[ 0\right] $ we find
\begin{equation}
\widetilde{Z}\left[ 0\right] =\exp \left\{ -\frac{L^{4}}{2}\int N\left(
k\right) \frac{d^{4}k}{\left( 2\pi \right) ^{4}}\ln \frac{k^{2}+m^{2}}{\pi }%
\right\} .  \label{zf}
\end{equation}%
In particular, we can write $\widetilde{Z}\left[ 0\right] =\exp \left(
-L^{4}<\rho >_{eff}\right) $, where $<\rho >_{eff}$ is the zero-point vacuum
energy density which for a particular topology $N\left( k\right) $ is%
\begin{equation}
<\rho >_{eff}=\frac{1}{2}\int N\left( k\right) \frac{d^{4}k}{\left( 2\pi
\right) ^{4}}\ln \frac{k^{2}+m^{2}}{\pi }.  \label{Lambd}
\end{equation}%
Thus we see whether the cosmological constant is finite or not depends on
the topological structure of the actual space. Now to account for all
possible virtual topologies (spacetime foam) and get the final expression
for the generating function $Z\left[ J\right] $ we have to sum over
topologies, i.e., possible values of $N\left( k\right) $ in accordance to (%
\ref{z}). For sure we may expect that all topologies which give infinite
values of $<\rho >_{eff}$ should be suppressed.

\section{Cutoff function in particle physics}

While the topology is fixed, $N\left( k\right) $ is an ordinary fixed
function\footnote{%
Actually $N\left( k\right) $ defines the whole topological class, while a
specific topology is fixed by a function $N\left( k,k^{\prime }\right) $.}.
Now we are ready to evaluate the cutoff for the particle physics in the case
when topology may fluctuate. In this case possible values of $N\left(
k\right) $ are $0$ and $1$. The partition function (\ref{zf}) has the
structure%
\[
\widetilde{Z}\left[ 0\right] =\prod\limits_{k}Z_{k}^{N\left( k\right) }
\]%
where $Z_{k}$ is given by the standard single-field expression $Z_{k}=\sqrt{%
\pi /\left( k^{2}+m^{2}\right) }$ and the sum over possible values $N\left(
k\right) $ gives%
\begin{equation}
Z=\sum_{topologies}\widetilde{Z}\left[ 0\right] =\prod\limits_{k}\left(
\sum_{N=0,1}Z_{k}^{N\left( k\right) }\right) =\prod\limits_{k}\left(
1+Z_{k}\right) ,  \label{prt}
\end{equation}%
while for the mean cutoff we find from (\ref{z1})%
\begin{equation}
\overline{N}\left( k\right) =\frac{Z_{k}}{\left( 1+Z_{k}\right) }.
\label{cutoff}
\end{equation}%
This expression straightforwardly generalizes on a multiplet of scalar
fields or a set of bosonic fields of an arbitrary spin which gives
\begin{equation}
\ln Z_{k}=\frac{1}{2}\sum_{\alpha }\ln \frac{\pi }{\left( k^{2}+m_{\alpha
}^{2}\right) },  \label{zb}
\end{equation}%
where the sum is taken over all fields and helicity states. The bias and the
cutoff for Fermi fields require a separate consideration which we consider
elsewhere\footnote{%
Standard fermionic fields have the negative energy in the ground state which
leads to the instability of fermionic fields with respect to topology
fluctuations. However, we point out that the action for fermionic fields has
no the classical limit and therefore it is defined up to a constant shift
(the cosmological constant term). Moreover, if we consider some coarse
graining in the phase space, then the difference between fermions and bosons
should disappear (upon the coarse graining, more than one fermion can occupy
the same quantum state). By other words the instability pointed out should
lead to a phase transition upon which fermionic excitations acquire positive
energy density in the ground state and become stable with respect to
topology fluctuations.}.

The remarkable property of the cutoff function is the explicit Lorentz
invariance (i.e., the function $\overline{N}\left( k\right) $ depends on the
momenta via the Lorentz invariant expression $k^{2}$). On the mas-shell $%
Z_{k}\rightarrow \infty $ and it reduces to $N\left( k\right) \rightarrow 1$
which reflects the fact that on the mas shell the space looks as $R^{4}$,
while at very small (planckian) scales $Z_{k}\ll 1$ it has the behavior $%
N\left( k\right) \sim 1/k^{g}\rightarrow 0$ as $k\rightarrow \infty $,
(where $g$ is the total number of degrees of freedom). Thus, as it was
expected \cite{wheeler} for sufficiently big number of fields $g$, $N\left(
k\right) $ provides indeed a Lorentz invariant cutoff which we discuss in
the next section.

\section{Finiteness of Feynman diagrams}

The generating functional $\widetilde{Z}\left[ J\right] $ leads to the
standard perturbation scheme (e.g., see the standard textbooks \cite{TB}).
New features however appear. As we can see from (\ref{fg}) and (\ref{gf})
the integration measure for every closed loop takes the form $N\left(
k\right) d^{4}k/(2\pi )^{4}$ and, therefore, every diagram will include the
factor $\left\langle N\left( k_{1}\right) N\left( k_{2}\right) ...N\left(
k_{n}\right) \right\rangle $ which in the first only approximation by
topology fluctuations can be replaced with the product $\overline{N}\left(
k_{1}\right) \overline{N}\left( k_{2}\right) ...\overline{N}\left(
k_{n}\right) $ where $\overline{N}\left( k\right) $ gives the cutoff which
is defined by (\ref{cutoff}). Thus every Feynman diagram acquires an
additional decomposition onto a series by topology fluctuations of the
cutoff function.

The contribution in the cutoff function $\overline{N}\left( k\right) $ comes
from all physical fundamental fields (\ref{zb}) and it is clear that all UV
divergencies are automatically regularized (e.g., if we account only for
gravitational $h_{\mu \nu }$, electromagnetic $A_{\nu }$, and weak $Z_{\nu }$%
, $W_{\nu }^{\pm }$ \ interactions, the number of degrees of freedom is $10$
and it defines the UV behavior $N\left( k\right) \sim 1/k^{10}$ as $%
k\rightarrow \infty $ which is already sufficient to regularize all
divergent diagrams\footnote{%
We point out that gauge fields have more components whose contribution to $%
Z_{k}$ depends on the choice of the gauge fixing. Therefore the exponent in $%
N\left( k\right) \sim 1/k^{g}$ may be even more than ten.}. In (\ref{cutoff}%
) the characteristic UV scale of the cutoff has the planckian order $%
Z_{k}\sim 1$, which means that $Z_{k}$ includes contribution of all fields
with mas less than planckian mas $m_{pl}$. This is not convenient for
practical computations; for the actual cutoff occurs for much lower
energies. To see this let us introduce the characteristic scale $k\sim \mu $
which has the sense of the laboratory scale from which we extrapolate our
laboratory coordinate system to very small distances (i.e. the actual scale
of the cutoff). From the analogy with the statistical physics such a scale
can be viewed as a specific chemical potential which corresponds to the
additional cosmological constant term to the action\footnote{%
We recall that when we consider interactions all constants acqire a
dependence on scales \cite{TB}.}, i.e., the redefinition of (\ref{Lambd}) as
\begin{equation}
<\rho >_{eff}=\frac{1}{2}\int N\left( k\right) \frac{d^{4}k}{\left( 2\pi
\right) ^{4}}\ln \frac{k^{2}+m^{2}}{\mu ^{2}}.  \label{l}
\end{equation}%
Then $Z_{k}$ modifies as $Z_{k}\rightarrow Z_{k}/Z_{\mu }$ and the cutoff
function (\ref{cutoff}) modifies as%
\begin{equation}
\overline{N}\left( k\right) =\frac{Z_{k}}{\left( Z_{\mu }+Z_{k}\right) }.
\end{equation}%
In such a form we may retain in $Z_{k}$ only the necessary (smallest) number
of fields with masses $m_{\alpha }<\mu $, while all more massive particles
give only a constant contribution to $Z_{k}\sim \mu /m$ and lead merely to a
renormalization of the scale $\mu $ itself. By other words we may suppose
that the contribution of the most heavy particles is already encoded in $\mu
$ (at least this allows also to account phenomenologically for all possible
new particles and fields which may be found in the future at extremely high
energies).

Thus the cutoff function acquires the structure
\begin{equation}
\overline{N}\left( k\right) =\frac{\mu ^{g}}{\left( \mu ^{g}+k^{2\alpha
_{0}}\left( k^{2}+m_{1}^{2}\right) ^{\alpha _{1}}\cdots \left(
k^{2}+m_{n}^{2}\right) ^{\alpha _{n}}\right) }  \label{cut}
\end{equation}%
where $m_{\alpha }<\mu $ and $g=\sum 2\alpha _{n}$ is the total number of
fields we have to retain.

The most divergent expressions in quantum field theory come from terms of
the type $\left\langle \left( \partial \phi \right) ^{2}\right\rangle $,
which in the momentum space have UV behavior\footnote{%
Actually the most divergent behavior will be given by $\sim p^{8}$, when
fluctuations in the cutoff function itself are taken into account, since the
Gaussian character of the distribution over $N\left( k\right) $ gives $%
\overline{\Delta N^{2}}\sim \overline{N}$. However such terms should be
treated in the complete analogy with the subsequent analysis.} $\sim p^{4}$.
We point out that $p^{4}$ gives also the highest rate of divergency in
quantum gravity as well, e.g. see Ref. \cite{DW}. As an example of such a
term we consider the cosmological constant (\ref{l}). Since all fields which
we retain in (\ref{cut}) give some contribution to the cosmological constant
$<\rho >_{eff}$ we sum (\ref{l}) over all fields which gives (upon simple
transformations)%
\begin{equation}
<\rho >_{eff}=\frac{\mu ^{4}}{\left( 16\pi ^{2}\right) }F\left( \alpha ,%
\widetilde{m}\right) ,
\end{equation}%
where%
\[
F\left( \alpha ,\widetilde{m}\right) =\int_{0}^{\infty }\frac{\ln \left(
x^{\alpha _{0}}\left( x+\widetilde{m}_{1}\right) ^{\alpha _{1}}\cdots \left(
x+\widetilde{m}_{n}\right) ^{\alpha _{n}}\right) }{\left( 1+x^{\alpha
_{0}}\left( x+\widetilde{m}_{1}\right) ^{\alpha _{1}}\cdots \left( x+%
\widetilde{m}_{n}\right) ^{\alpha _{n}}\right) }xdx
\]%
and $\widetilde{m}_{i}=m_{i}^{2}/\mu ^{2}$. This expression is finite for $%
\sum 2\alpha _{n}>4$ (i.e., we have to retain at least five field degrees of
freedom). In the case when $\widetilde{m}_{i}=0$ it gives%
\[
F\left( \alpha ,0\right) =-\frac{\pi ^{2}}{\alpha _{0}}\frac{\cos \left(
2\pi /\alpha _{0}\right) }{\sin ^{2}\left( 2\pi /\alpha _{0}\right) }.
\]

Next "dangerous" terms are given by $\left\langle \phi ^{2}\right\rangle $
which define the renormalization of the mas. We evaluate it for $\lambda
\phi ^{4}$ \cite{TB} which in the first order by $\lambda $ gives the
correction to the mas (the so-called "tadpole" diagram)
\begin{equation}
\delta m^{2}=\Sigma \left( p\right) =\frac{\lambda }{2}\int N\left( k\right)
\frac{d^{4}k}{\left( 2\pi \right) ^{4}}\frac{1}{k^{2}+m^{2}}  \label{tad}
\end{equation}%
which gives
\[
\Sigma \left( p\right) =\Sigma \left( 0\right) =\frac{\lambda }{32\pi ^{2}}%
\mu ^{2}G\left( \alpha ,\widetilde{m}\right) ,
\]%
where
\[
G\left( \alpha ,\widetilde{m}\right) =\int_{0}^{\infty }\frac{xdx}{\left( x+%
\widetilde{m}\right) \left( 1+x^{\alpha _{0}}\left( x+\widetilde{m}%
_{1}\right) ^{\alpha _{1}}\cdots \left( x+\widetilde{m}_{n}\right) ^{\alpha
_{n}}\right) }
\]%
which is already finite for $\sum 2\alpha _{n}>2$. In the massless case it
gives
\[
G\left( \alpha ,0\right) =\frac{1}{\alpha _{0}}\Gamma \left( 1/\alpha
_{0}\right) \Gamma \left( 1-1/\alpha _{0}\right) .
\]

In this manner we see that all divergencies in Feynman diagrams disappear
when the\ contribution of a proper number of fields in the cutoff function
is taken into account. It is quite clear that this result is valid almost in
all theories (whose dynamical equations do not include too high derivatives
of fields which in general lead to $p^{n}$ divergencies) and it seems to
remain true in general relativity (GR) as well \cite{DW}. However unlike
gauge fields (which are proved to be renormalizable) GR represents formally
non-renormalizable theory\footnote{%
There is only a small chance that due to the entanglement in complex
diagrams divergencies may remain.} and therefore it requires the more
complete and rigorous proof which we leave for the future research.

\section{Cutoff function on the universal covering}

In observational cosmology when we look at the sky we always use the
coordinate system which corresponds to the universal covering. Therefore, in
solving astrophysical problems (quantum origin of density perturbations,
quantum cosmology, etc.) we have to use the representation in which the
actual space is described by the universal covering. We recall that in
general the universal covering requires the introducing of a curved
background and, therefore, the results of the present section have only a
preliminary character.

In the present section we evaluate the astrophysical cutoff function as
well. In this case the number of fields takes the values $N\left( k\right)
=0,1,2,...$ and (\ref{prt}) becomes
\begin{equation}
Z=\sum_{topologies}\widetilde{Z}\left[ 0\right] =\prod\limits_{k}\left(
\sum_{N=0}^{\infty }\frac{Z_{k}^{N\left( k\right) }}{N\left( k\right) !}%
\right) =\exp \left( L^{4}\int Z_{k}\frac{d^{4}k}{\left( 2\pi \right) ^{4}}%
\right) ,  \label{zz}
\end{equation}%
where we have accounted for the fact that permutations of fields at the same
$k$ gives the same quantum state (i.e., the identity of fields which gives
the factor $1/N!$). Then for the mean cutoff we find from (\ref{z1})
\begin{equation}
\overline{N}\left( k\right) =Z_{k}\ .  \label{ctf}
\end{equation}%
Thus (\ref{cutoff}) and (\ref{ctf}) define the relation between the bias
(cutoffs) in the two different representations for the same physical space.

The analogy with the statistical physics shows that (\ref{zz}) (\ref{ctf})
correspond to the classical (or the Boltzmann) statistics. As it was
discussed in the introduction such statistics corresponds to the so-called
diffused fields \cite{dif}. However quantum topology should introduce some
additional statistics between fields \cite{K99} which corresponds to third
quantization and which we consider in what follows\footnote{%
Such correlations may be important in investigating corrections to the mean
values of the type $\left\langle N\left( k_{1}\right) N\left( k_{2}\right)
...N\left( k_{n}\right) \right\rangle $ which appear in Feynman diagrams.}.

Consider first the density of fields in the configuration space (i.e., the
space of fields)%
\[
N\left[ k,\phi \right] =\sum_{j}\delta \left( \phi -\phi _{k}^{j}\right)
\]%
so that the number of fields is merely
\[
N\left( k\right) =\int N\left[ k,\phi \right] d\phi .
\]%
Then the action (\ref{act}) can be rewritten as%
\[
S=\frac{L^{4}}{2}\int N\left[ k,\phi \right] \left( k^{2}+m^{2}\right)
\left\vert \phi \right\vert ^{2}D\phi \frac{d^{4}k}{\left( 2\pi \right) ^{4}}
\]%
which represents the functional of $N\left[ k,\phi \right] $. Thus, the
partition function can be presented as
\[
Z=\sum_{N\left[ k,\phi \right] }\exp \left\{ -S\left( N\left[ k,\phi \right]
\right) \right\} .
\]%
Here the sum over $N\left[ k,\phi \right] $ includes, in fact, both the sum
over topologies and configuration variables. The further depends on the
statistics of fields assumed (which is not the same as the statistics of
particles, e.g., see Refs. \cite{K99,K03}). If we accept the Fermi
statistics (i.e., numbers $N\left[ k,\phi \right] =0,1$) then such scalar
particles will obey the so-called para-Bose statistics \cite{K03}. The
choice should be made from experiment (though there may be some theoretical
reasoning for a particular choice). In both cases we find for the mean
density%
\[
\left\langle N\left[ k,\phi \right] \right\rangle =\left[ \exp \left( \frac{1%
}{2}\left( k^{2}+m^{2}\right) \left\vert \phi \right\vert ^{2}\right) \pm 1%
\right] ^{-1}
\]%
and for the cutoff function we find the same expression (\ref{ctf}) with an
additional multiplier
\[
\overline{N}\left( k\right) =C_{\pm }Z_{k}^{g}
\]%
where the multiplier is given by ($g$ is the number of components of the
scalar field $\phi $)
\[
C_{\pm }=\frac{1}{\pi ^{g/2}}\int \frac{d^{g}\phi }{\exp \left( \frac{1}{2}%
\left\vert \phi \right\vert ^{2}\right) \pm 1}.
\]

\section{Conclusions}

In conclusion we briefly repeat basic results. First of all we have
explicitly demonstrated that spacetime foam provides quantum fields with a
cutoff. The form of the cutoff is fixed by the field theory itself and it
does not introduce additional parameters. It depends only on the standard
set of naked parameters related to fields. It does also depend on the
representation of the physical space used. We have to used the two types of
different representations depending on the problem under consideration. In
particle physics we extrapolate the laboratory coordinate system to
extremely small scales and, therefore, we should use the so-called standard
representation (the image method) which gives (\ref{cut}) for the cutoff. In
astrophysics however we always have deal with the universal covering and the
cutoff becomes (\ref{ctf}). Since we considered quantum topology
fluctuations around the flat space, the cutoff has the Lorentz invariant
form. This is always justified for particle physics, while in the
astrophysical picture our results carry rather a preliminary character; for
rigorous consideration requires a curved background. In the present Letter
our consideration has a simplified character, i.e., a set of scalar fields.
However it is clear that all the results can be straightforwardly extended
to any non-linear field theory. In particular, the cutoff suggested
automatically regularizes divergencies in quantum fields and, therefore, we
can expect that general relativity represents in fact a renormalizable
theory.

We also demonstrated that every Feynman diagram acquires an additional
decomposition onto a series by topology fluctuations in the cutoff function
which may lead to some new phenomena.

The cutoff function has the meaning of the topological bias of point sources
which displays the discrepancy between the visual and the actual spaces. In
astrophysics such a discrepancy is observed as the Dark Matter phenomenon
\cite{KT07,K06}. Analogous phenomena are widely known in particle physics
which represent "Dark Charges" of all sorts. Those are not more than the
standard (phenomenological) Higgs fields \cite{Higgs}. Therefore, we expect
that quantum gravity provides the unique tool to fix all constants of nature
(the lambda term, mas spectrum, charge values, etc.). However the
self-consistent evaluation of such parameters requires considering the
complete theory which is to be developed.

\section{Acknowledgment}

We acknowledge D. Turaev for useful discussions. For A.A. K this research
was supported in part by the joint Russian-Israeli grant 06-01-72023.

\end{document}